\documentclass[useAMS,usenatbib]{mn2e}

\usepackage{defs}
\usepackage{psfig}

\title[Investigating the Far-IR/Radio Correlation of Star Forming Galaxies to $z=3$]{Investigating the Far-IR/Radio Correlation of Star Forming Galaxies to $z=3$}
\author[N. Seymour et al.]{N. Seymour$^{1,2}$\thanks{E-mail: nps@mssl.ucl.ac.uk}, M. Huynh$^{3}$, T. Dwelly$^{4}$, M. Symeonidis$^{1}$, A. Hopkins$^{5}$, \and I.M. M$^{\rm c}$Hardy$^{4}$, M. Page$^{1}$ \& G. Rieke$^{6}$\\ 
$^{1}$Mullard Space Science Laboratory, UCL, Holmbury St Mary, Dorking, Surrey RH5 6NT.\\
$^{2}$Spitzer Science Center, Caltech, 1200 East California Boulevard, Pasadena, CA 91125, USA.\\
$^{3}$Infrared Processingand Analysis Center, MS220-6, California Institute of Technology, Pasadena, CA, 91125, USA.\\
$^{4}$School of Physics \& Astronomy, University of Southampton, Highfield, Southampton SO17 1BJ.\\
$^{5}$Anglo-Australian Observatory, PO Box 296 Epping, NSW 1710, Australia.\\
$^{6}$Steward Observatory, Tucson, USA.}

\begin{document}

\date{IN PRESS}

\pagerange{\pageref{firstpage}--\pageref{lastpage}} \pubyear{2009}

\maketitle

\label{firstpage}

\begin{abstract}
In order to examine the far-IR/radio correlation at high redshift we have 
studied the {\it Spitzer} $70\,\um$/far-infrared (far-IR) properties 
of sub-mJy radio sources from the $13^{\rm H}$ {\it XMM-Newton/Chandra} 
Deep Field by redshift and galaxy type: active 
galactic nucleus (AGN) or star forming galaxy (SFG). We directly detect 
$70\,\um$ counterparts (at $>3\,\sigma$ significance) for $22.5\%$ (92/408) 
of the radio sources, while for the rest we perform stacking analysis by 
redshift and galaxy type.
For the sources detected at $70\,\um$ we find that the 
median and scatter of the observed flux density ratio, q$_{\rm 70}$, are 
similar to 
previous results in the literature, but with a slight decrease in q$_{\rm 70}$
towards higher redshifts. Of the radio sources detected at $70\,\um$ 
8/92 were already 
classified as AGN, but two of which maybe SFGs.
For the stacked sources we obtain a detection for the 
SFGs at every redshift bin which implies they have mean flux densities a 
factor $\sim5$ below 
the original $70\,\um$ detection limit. For the stacked AGN
we obtain a detection only in our highest redshift bin ($1\le z\le 5$)
where we may be sampling hot dust associated with 
the AGN at rest-frame $12-35\,\um$.
The combined observed mean value of q$_{\rm 70}$ for the SFGs (detected
and non-detected at $70\,\um$) decreases gradually with redshift,
consistent with tracks derived from empirical spectral energy
distributions (SEDs) of local SFGs.
Upon closer inspection and when comparing with tracks of appropriate
luminosity, the values of q$_{\rm 70}$ broadly agree at low redshift.
However, at $z\sim 1$, the observed q$_{\rm 70}$ (for ULIRGs) is
$2\,\sigma$ below the value seen for local ULIRGs tracks, implying a
difference in the SED between local and $z\sim 1$ ULIRGs.
At higher redshifts, the convergence of the tracks and the higher
uncertainties in q$_{\rm 70}$ do not allow us to determine if such a
difference persists.
\end{abstract}

\begin{keywords}
galaxies: evolution, starburst, radio continuum: galaxies, infrared: galaxies 
\end{keywords}

\section{Introduction}
The tight correlation over many orders of magnitude between the far-infrared 
(far-IR) and radio luminosity of star forming galaxies (SFGs) has been well studied 
in the local Universe \citep[e.g.][]{Helou:85, Condon:91, Yun:01}. The radio 
emission of normal galaxies is dominated by 
synchrotron radiation from relativistic electrons and free-free emission from 
HII regions \citep{Condon:92}. Both these mechanisms are related to the presence
of young, massive ($M>8M_\odot$) stars; radio non-thermal emission arises from 
electrons accelerated by supernovae from these stars and thermal emission from 
ionized HII regions. The mid- and far-IR emission arises from dust absorption 
and subsequent re-radiation of UV/optical light.
Deviations from this correlation are seen within galaxies 
\citep{Hughes:06, Murphy:06}, but these probably derive from the longer 
mean free path of the relativistic electrons compared to the dust 
heating UV photons and the scatter in the relation within local galaxies is 
similar to that 
seen between local galaxies \citep[e.g. ][]{Yun:01}. Some authors 
\citep{Hughes:06,Murphy:06} have suggested that the far-IR/radio 
correlation may also have some dependence on star formation rate (SFR) and 
SFR density within a galaxy.

The far-IR/radio correlation is important as it is used to define the radio 
luminosity/SFR 
relation \citep[e.g.][]{Bell:03} and can also allow the selection 
of radio-loud active galactic nuclei (AGN) \citep[e.g.][]{Donley:05}. 
Hence, determining whether this relation deviates significantly or becomes less tight 
at higher redshifts and higher luminosities is of great consequence. 
In particular the interpretation of data from the next generation of radio 
surveys, which will find Milky Way-like star forming galaxies out to high redshifts, 
will rely on the application of the radio luminosity/SFR relation.
In the distant Universe the relation between the mid-IR and radio luminosities 
has been shown to approximately hold to 
$z\sim 1$ \citep{Garrett:02, Appleton:04} based on {\it ISO} and {\it Spitzer} 
observations, although there are some disagreements about the proportionality 
constant. 
At $z\ge 1$ using millimetre and sub-millimetre observations 
\citet{Kovacs:06} and \citet{Vlahakis:07} find evidence of a decrease of 
$\sim0.2\,$dex in the far-IR/radio ratio compared to the 
local value, but it was shown in \citet[][hereafter S08]{Seymour:08} that 
the relation does approximately hold by comparing AGN-free distant and local 
samples of powerful starbursts at the same rest-frame far-IR wavelengths. 
However the S08 study was small consisting of eleven sources at $1.5\ge z\ge 3$.

Several different mechanisms could lead to departures from the radio/far-IR 
correlation at high redshift. Deviations at low luminosities have been noticed
\citep{Bell:03}, but these occur in low SFR sources (SFR$\,\le 3\,$M$_\odot$yr$^{-1}$) 
which cannot be detected to any great distance in current radio surveys. Modeling 
of the radio/far-IR relation in dusty starbursts has shown that the correlation is
a natural result in any starburst if synchrotron emission dominates inverse 
Compton, and the electron cooling time is shorter than the analogous fading time 
of the 
supernovae rate \citep{Bressan:02}. However, deviations may be possible in both 
the early phase of a starburst, when the radio thermal component dominates the 
non-thermal component, and in the post-starburst phase, when the bulk of the 
non-thermal component originates from less massive stars. 
Other potential mechanisms that could lead to deviations from the
relation include: (1) evolution in metallicity and dust properties,
with accompanying changes
in the far-IR spectral energy distributions (SEDs); 
(2) evolution in magnetic field properties;
(3) quenching from the Cosmic Microwave Background; or
(4) the effects of hot intracluster gas in
dense environments \citep[e.g.][]{Miller:01}.

The most sensitive infrared probe of high redshift star formation
with current facilities is the $24\,\um$ band of the Multiband Imaging 
Photometer for Spitzer (MIPS) instrument 
\citep{Rieke:04} on board the {\it Spitzer Space Telescope} \citep{Werner:04}. 
The cross-correlation of radio and $24\,\um$ data from many deep surveys 
has led to several evaluations of a mid-IR/radio relation 
\citep{Boyle:07,Beswick:08,Garn:09,Ibar:08}. Some of these authors 
\citep{Boyle:07, Beswick:08} observe a change in the mid-IR/radio 
ratio at faint radio and/or $24\,\um$ flux densities, but as these authors 
discuss there are different populations being analysed depending on the 
selection of the parent population. Also, studies using $24\,\um$ observations 
suffer from the shift of this band with redshift to shorter wavelengths where 
it  no longer tracks the cool dust directly associated with star formation 
and into a regime where features due to polycyclic aromatic hydrocarbons 
(PAHs) may dominate.  Furthermore, it is possible that different 
processes (star formation or AGN) dominate in each of the radio and $24\,\um$ 
bands.

We aim to overcome some of these difficulties by using a sample of sub-mJy 
radio sources that have been categorised by the physical process that 
dominates their radio emission: star formation or AGN activity. In 
S08 we used several radio-related 
discriminators (radio morphology, radio spectral index, observed $24\,\um$/radio 
and $K-$band/radio flux density ratios) to separate these two populations 
in a very deep $1.4\,$GHz survey and confirmed the expected result  
\citep[from modeling of the radio source counts ---][]{RowanRobinson:91, Hopkins:04} 
that the SFG population dominates at the 
faintest radio flux densities, $S_{\rm 1.4GHz}\le 0.1\,$mJy, 
but with a significant, $\sim\frac{1}{3}$, contribution from AGN. 

In this paper we examine the far-IR/radio correlation of the sub-mJy 
radio population by galaxy type and redshift by studying their 
$70\,\um$ emission. Studying the emission at $70\,\um$ has several 
advantages over that at $24\,\um$. This far-IR band is more
comparable to the infrared wavelengths used in the early studies of this
correlation \citep[i.e., using the IRAS $60$ and $100\,\um$ bands;][]{Helou:85}.
The $70\,\um$ band also samples the SED of local (U)LIRGs closer to its peak 
($60\,\um\le\lambda_{\rm rest}\le 120\,\um$) 
than the $24\,\um$ band and is a more direct measure of the total IR 
luminosity and hence SFR. Furthermore, given the redshift range
of the sample examined here, an analysis at $24\,\um$ would probe rest-frame
wavelengths $7\,\um\le \lambda_{\rm rest}\le 24\,\um$,
well away from the cold dust related to the ongoing star formation, and 
into a regime containing uncertain tracers of star formation (e.g. PAHs) as 
well as complications due to silicate absorption/emission features and possible 
AGN contamination of unknown strength. 
The $70\,\um$ band has a few disadvantages: its relative shallowness compared 
to equivalent surveys at $24\,\um$ and the potential confusion at faint flux 
densities due to the low resolution at $70\,\um$. We can overcome
the first of these issues through the use of stacking techniques on 
sources not detected at $70\,\um$. The latter problem is mostly overcome by the 
accurate (sub-arcsec) positions of our radio sources, the parent population 
we are investigating.

We have chosen to adopt in this paper a similar philosophy to that used in S08,
where we make no assumptions about the SED of the sources we observe,
but use {\it observed} flux density ratios as a function of redshift
at all times, and then compare these to models. This approach is the most 
appropriate since at each redshift slice we are looking at sources with different
luminosities which may not be directly comparable to each other. 
We present our data analysis and cross-correlation in \S 2 and in \S 3 we 
present the results of our cross-correlation and stacking analysis. We discuss 
our results in \S 4 and present our conclusions in \S 5.
Throughout we use a concordance model of Universe 
expansion, $\Omega_M = 1 - \Omega_{\Lambda} = 0.3$, $\Omega_0 = 1$, and 
$H_0 = 70\, \kmpspMpc$ \citep{Spergel:03}.

\section{Data Analysis}

\subsection{Radio Data and Separation by Galaxy Type}

The radio sample used in this work was originally presented in 
\citet[][hereafter S04]{Seymour:04} which described our deep Very Large 
Array (VLA) $1.4\,$GHz observations of 
the $13^{\rm H}$ {\it XMM/Chandra} Deep Survey Field \citep{McHardy:98a}. 
We obtained a rms noise of $7.5\,\uJy$  at the center of the single 
VLA pointing and found 449 sources at $4\,\sigma$ significance out to a 
diameter of $30\,$arcmin. Over the last decade we have obtained deep follow-up
optical, near-IR and {\it Spitzer} imaging. 
We have obtained spectra for 163/449 radio sources and 
these spectra were used to calibrate our 14-band optical/near-/mid-IR 
photometric redshifts for the radio sources (see Dwelly et al. in prep. for a 
summary). Our spectroscopy revealed one radio source to 
be a star which we remove from further consideration. Only sources that were 
detected in at least 4\,bands (236 sources) have photometric redshifts.
Hence, our parent sample consists of 448 radio sources of which 162 have 
spectroscopic redshifts, 236 have photometric redshifts and 50 sources have 
unknown redshifts.

In S08 we described the separation of our parent sample into galaxy type: AGN 
and SFG. In that paper we made the assumption that one of these processes 
dominated the radio emission and that we could use diagnostics directly 
related to the radio emission to separate these radio sources. These 
empirical diagnostics were radio morphology (from combined MERLIN/VLA 
observations), radio spectral index (from combined VLA $1.4/4.8\,$GHz
data), and two observed flux density ratios plotted as function of redshift:
$24\,\um$/radio and radio/$K-$band. Our analysis in S08 found 269/449 faint 
$1.4\,$GHz sources to be SFGs and the remainder to be AGN.

Our radio catalogue is not complete to the 
$30\,\uJy$ detection limit due to radio 
instrumental effects. The attenuation of the primary beam
of the VLA away from the pointing center is by far the strongest effect and
causes a decrease in the sensitivity of the survey by a factor of $\sim2$
at the edge of the field of view. In S04 we calculated 
correction factors due to this effect (as well as other more minor ones)
and used these correction factors in determining the radio source counts;
each source was given a weight corresponding to its detection plus 
a factor representing all the sources at the same flux density, but not 
detected due to the effective decrease in the sky area at faint flux 
densities. These correction factors were also used in determining the 
comoving SFR density  history in S08 and we use these same factors again 
here. These factors only apply to sources 
below $100\,\uJy$ and are generally low, $\le 2$ for 
$50\,\uJy<S_{1.4GHz}<100\,\uJy$, but do reach values $\ge 2$
at $S_{1.4GHz}<50\,\uJy$. We discuss the impact of the use of these 
correction factors in section~\ref{sec.bias}.

\begin{figure*}
\centerline{
\psfig{file=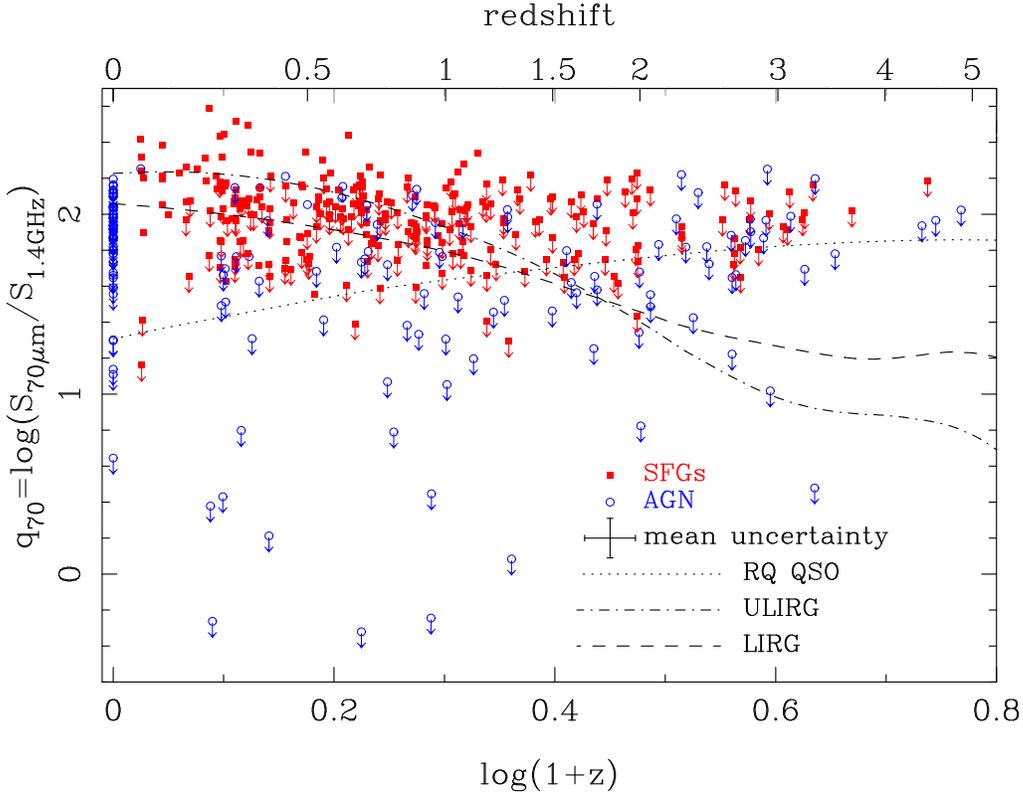,width=13.5cm,angle=270}
}
 \caption{The {\it observed} far-IR/radio flux density ratio 
(q$_{\rm 70}=\log[S_{70\um}/S_{\rm 1.4GHz}]$) of our radio-selected population
plotted as a function of redshift; sources with unknown 
redshifts are plotted at $z=0$. Sources are separated by galaxy type 
(SFGs: red squares, AGN: blue open circles) and upper limits represent 
non-detections at $70\,\um$ at a $3\,\sigma$ limit of 6\,mJy. 
Average individual uncertainties are indicated although we note the 
uncertainty in the redshift is considerably 
less for the $\sim\frac{1}{3}$ sources with spectroscopic redshifts.
The tracks of local (Ultra) Luminous InfraRed Galaxies, (U)LIRGs (derived from 
empirical observations, Rieke et al. 2009) are shown as reference.
The ULIRG ($\log(L_{\rm IR}/L_\odot)\ge 12$) has a higher SFR than the LIRG 
($\log(L_{\rm IR}/L_\odot)\ge 11$) by a factor of ten and has a higher 
observed value of q$_{\rm 70}$ until $z\sim2.5$. We also show the track of a 
radio-quiet QSO from Elvis et al. (1994). }
\label{fig:q70raw}
\end{figure*}

\subsection{{\it Spitzer} MIPS $70\,\um$ Data}

The MIPS observations of the $13^{\rm H}$ field consist of a deep scan map
using all three bands ($24$, $70$ and $160\,\um$) which covers
approximately $0.5\deg\times 1\deg$. These observations were obtained
in July 2005 as part of MIPS instrument team GTO time (PI G. Rieke, 
programme identification number 81). Our 1.4\,GHz radio observations 
cover one end of this MIPS map.
The $70\,\um$ Basic Calibrated Data (BCDs) from the Spitzer 
Science Center were processed offline using 
the Germanium Reprocessing Tools ({\tt GERT}\footnote{Available from 
{\tt http://ssc.spitzer.caltech.edu/postbcd/} }), 
following the filtering techniques 
adopted for the extragalactic First Look Survey \citep[xFLS;][]{Frayer:06}. 
In particular, negative sidelobes near bright sources were removed using a 
combination of high pass time median and column filtering with the bright 
sources masked. The BCDs were then mosaiced in the standard fashion with 
{\tt MOPEX$^1$} and the rms noise in the final $70\,\um$ map was $\sim 2\,$mJy
per beam. A list of sources was produced 
with the {\tt APEX}$^1$ software, which performs robust PSF fitting, down 
to $3\,\sigma$ ($6\,$mJy). Flux densities were divided by a factor of 0.918 to 
colour correct for a $\nu\times F_\nu=\,$constant SED (the most appropriate 
for the sources we are detecting) as the MIPS absolute flux 
calibration references a $10^4\,$K blackbody. The source list is incomplete 
at the $3\,\sigma$ level of significance and may not be
free from spurious sources, but spurious sources are not a concern for our 
approach, where we simply intend to measure the $70\,\um$ flux density 
of each member of our parent radio sample. We note that the probability 
of an individual $70\,\um$ source being matched by chance with a radio source  
is low ($\le 1.5\,\%$).

Out of the 448 radio sources in the parent sample, 40 were not covered by the 
$70\,\um$ image. From the remaining 408 sources we find 92 with $70\,\um$ 
counterparts within $5''$ from our $\ge 3\,\sigma$ catalogue. These radio 
sources with $70\,\um$ counterparts all have redshifts and include eight
AGN and 84 SFGs. Hence, there remain 316 radio sources with only 
$70\,\um$ upper limits. 

\subsection{Stacking of $70\,\um$ Data}

To push our analysis to fainter $70\,\um$ flux densities, we made stacked 
postage stamp cutouts of the $70\,\um$ image at the locations 
of radio sources that were not matched to sources in the $70\,\um$ catalogue.
Each cutout was $128\,\times 128\,$arcsec ($32\times 32\,$pixels after 
resampling to $4\,$arcsec pixels in the mosaicing) in size.
These cutouts were combined using a weighted mean (where the 
weight is the the inverse of the square of the local rms). 

\begin{figure}
\centerline{
\psfig{file=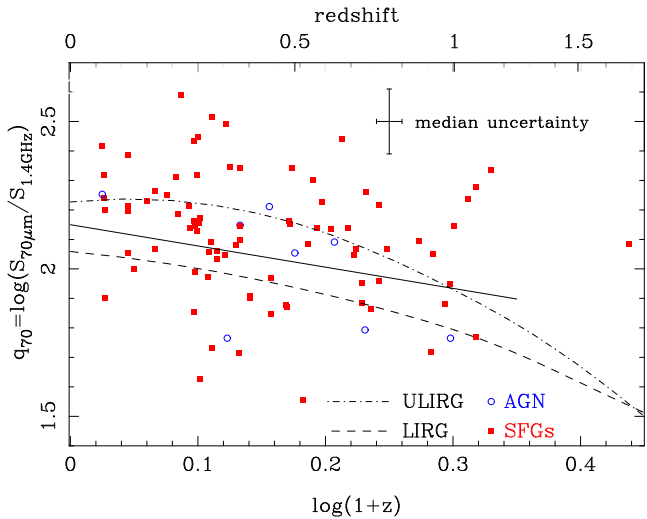,width=8.5cm,angle=270}
}
 \caption{The observed q$_{\rm 70}$ as function of redshift for 
just the radio sources detected at $70\,\um$ SFGs: red squares, AGN: blue 
open circles).
The solid line indicates the best fit linear regression to these data (SFGs 
and AGN). We 
discuss the $70\,\um$ detection of eight AGN in the text. }
\label{fig:q70raw2}
\end{figure}

The stacking was done in several bins of galaxy type (SFG, AGN)
and redshift. In S08 we found redshifts for all our SFGs so the sources
with unknown redshifts comprise one bin at $z=0$ of 47 AGN which are too 
faint in the optical and near-IR for photometric redshifts to be determined
(none of which are detected at $70\,\um$). 
The bins were initially chosen to be of equal size in $\log(1+z)$,
except some bins were amalgamated to ensure each bin had at 
least 30 sources: the two highest redshift SFG bins were combined and 
the five AGN bins were combined into two. 
Such amalgamation is necessary in order to obtain a significant improvement, 
i.e. $\ge\sqrt{30}$, in the sensitivity compared to the unstacked image. 
Similar to  \citet{Huynh:07b}, offset stacked 
images with the same number of sources per bin were generated by randomly 
choosing a nearby position  ($<64''$ or $<3.5\,$FWHM) in the $70\,\um$ image 
for each stacked source. Two hundred randomly offset stacks were generated, 
and the uncertainty in the stacked flux density is taken to be the 
standard deviation of these 200 measured values.

\section{Results} 

\begin{figure*}
\centerline{
  \psfig{file=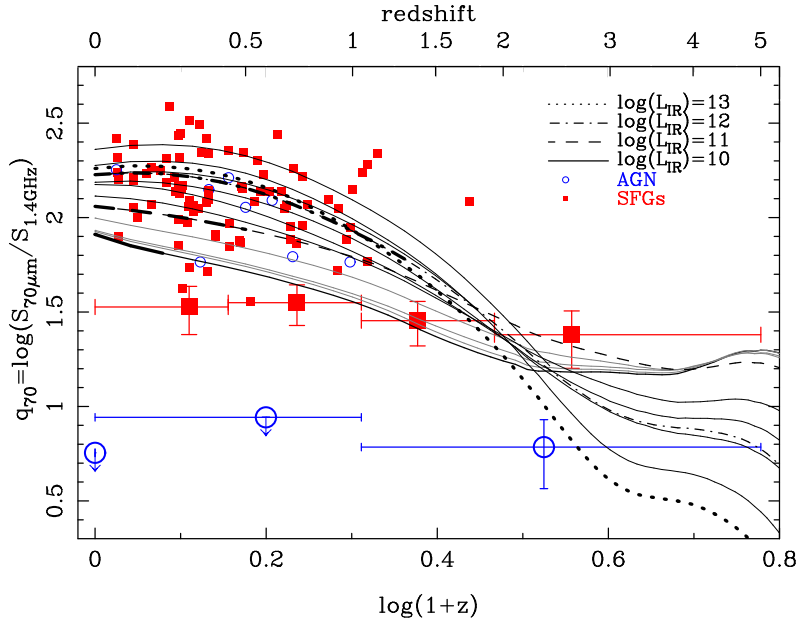,width=10.5cm,angle=270}
}	
\caption{The observed value of q$_{\rm 70}$ plotted as a function of
redshift and source type (SFGs: red squares, AGN: open blue circles)
for $70\,\um$ detected sources and stacked non-detections at $70\,\um$. 
Overlaid are star forming tracks covering a range of IR luminosities from 
\citet{Rieke:09} at luminosities indicated in the figure including those in Fig 1.
These lines become thinner at the redshift a template of a given luminosity
becomes undetectable in our radio survey. The thin grey lines represent tracks 
from templates at intermediate luminosities with interval of $0.25\,$dex.
The mean flux density ratios of sources 
undetected at $70\,\um$, derived from the stacked $70\,\um$ flux densities,
are indicated by larger symbols with error bars (indicating 
the width of the redshift bin and the uncertainty in q$_{\rm 70}$).
The SFGs have a detected flux density in each redshift bin and hence a 
determinable value of q$_{\rm 70}$ whereas AGN 
are only detected in the highest redshift bin.}
\label{fig:q70stackall}
\end{figure*}

\subsection{Observed q$_{\rm 70}$ for the Whole Sample}

The IR/radio correlation (at mid- or far-IR wavelengths) 
is commonly defined as q$_{\rm IR}=\log[S_{\rm IR}/S_{\rm 1.4GHz}]$ 
\citep[e.g.][]{Yun:01,Appleton:04, Beswick:08, Ibar:08}. We examine 
our {\em observed} values of q$_{\rm 70}$ in Fig.~\ref{fig:q70raw} as a function 
of redshift for all sources (including upper limits for the radio sources not 
detected at $70\,\um$). Fig.~\ref{fig:q70raw} figure 
shows a wide scatter for the whole parent population, 
although most values are upper limits due to non-detection at $70\,\um$. 
The sources with very low q$_{\rm 70}$ upper limits, as seen for many of the AGN, are 
typically the brightest radio sources in our survey.
As our radio survey is flux density limited, and hence suffers from some 
redshift/luminosity degeneracy, it is not appropriate to fit
to q$_{\rm 70}$ as a function of redshift without accounting for the different 
luminosity ranges probed at each redshift. We choose to plot tracks calculated
from templates of local galaxies, but shifted to the corresponding redshift 
and convolved with the appropriate band-passes, 
as a direct comparison to our {\em observed} values of q$_{\rm 70}$.
We overlay tracks derived from empirical local LIRG and ULIRG SEDs 
\citep{Rieke:09} and an unobscured radio-quiet AGN SED \citep{Elvis:94}. 
The starburst templates are derived from the mean of many local galaxies of 
the appropriate luminosity. We note an empirical M82 track would lie 
marginally below the LIRG track presented here. 

\subsection{Observed q$_{\rm 70}$ for the Detected Sample}

The radio sources detected at $70\,\um$ all have values of q$_{\rm 70}>1.5$ 
(Fig.~\ref{fig:q70raw2}) mainly due to the sensitivity limit of the $70\,\um$ 
observations. We calculate the median value of q$_{\rm 70}$ for 
this subsample in order to compare to earlier work.
We find a value of q$_{\rm 70}=2.13\pm0.24$ \citep[using a 
biweight estimator,][]{Beers:90}, very similar to that found by 
\citet[][]{Appleton:04}: q$_{\rm 70}=2.16\pm0.17$ (covering a similar redshift 
range). Additionally we derive values separately for each galaxy type 
obtaining a similar value for the 84 SFGs  (q$_{\rm 70}=2.13\pm0.24$) 
and a slightly lower value for the eight AGN
(q$_{\rm 70}=2.02\pm0.22$). While SFG and AGN results are marginally 
statistically different the similarity in values is probably due to the depth 
of the $70\,\um$ data not being able to detect sources, regardless of
type, with a low value of q$_{\rm 70}$. Naturally, the values of observed 
median q$_{\rm 70}$ for SFG and AGN are only upper 
limits for the radio selected sample as a whole and the true scatter is 
certainly greater. We also fit these data against 
$\log(1+z)$ by a simple linear regression as this measure of distance more 
closely follows look-back time. We find for the whole sample and by galaxy type:

\begin{equation}
  {\rm q}_{70}=2.15\pm0.10-0.72\pm0.29\times \log(1+z)~~~~(all)
\end{equation}			                      
\begin{equation}		                      
  {\rm q}_{70}=2.14\pm0.10-0.75\pm0.32\times \log(1+z)~~~~(SFGs)
\end{equation}			                      
\begin{equation}		                      
  {\rm q}_{70}=2.31\pm0.21-1.59\pm0.52\times \log(1+z)~~~~(AGN)
\end{equation}

We observe a decrease in q$_{\rm 70}$ with redshift for the whole of the $70\,\um$ 
detected sample which is principally due to sources with very high values of 
q$_{\rm 70}$ being found at low redshift, but which become much rarer at higher 
redshifts. This fit is shown in Fig~\ref{fig:q70raw2}.
The sample fitted here is naturally biased against low values of 
q$_{\rm 70}$ due to the $70\,\um$ detection limit. We find that the SFGs show a
similar trend to the total population, but although the AGN have a 
steeper slope they are almost consistent within the uncertainties and the 
slope is largely dependent on the lowest and highest redshift sources.
All of these slopes are consistent with the decrease 
in observed q$_{\rm 70}$ seen in local templates redshifted to earlier epochs, 
e.g. as seen in Figure~\ref{fig:q70stackall}.

\begin{figure*}
\centerline{
  \psfig{file=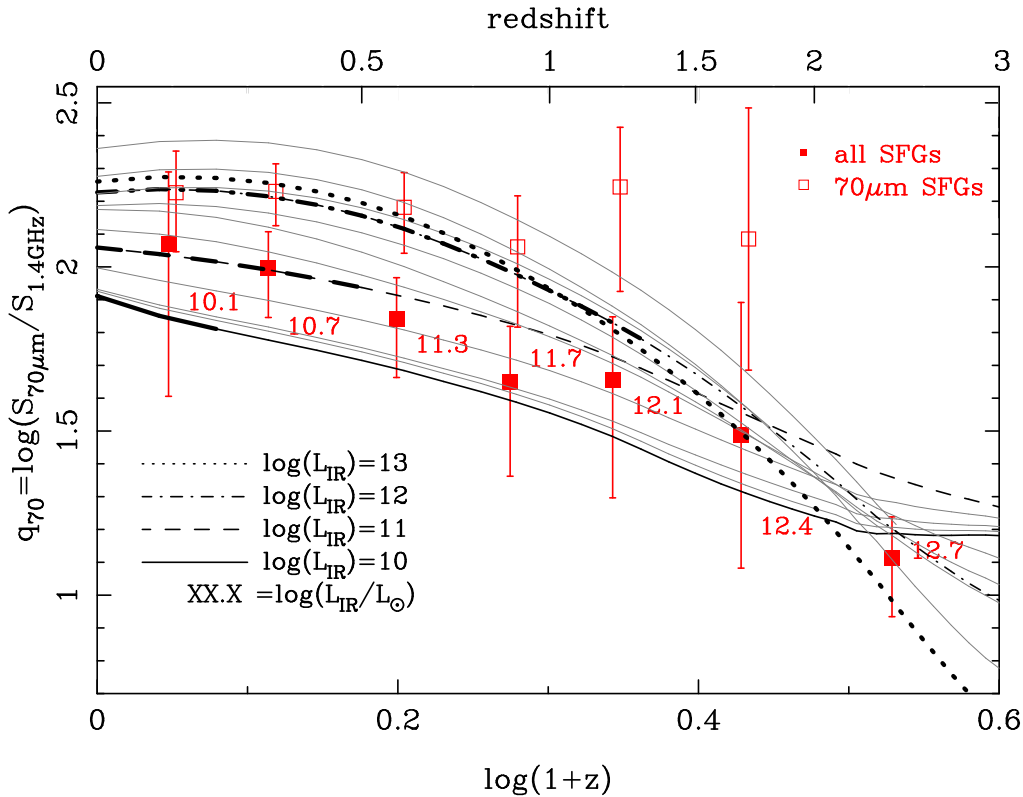,width=13.5cm,angle=270}
}	
 \caption{The mean value q$_{\rm 70}$ for all SFGs plotted as a function of 
redshift where we have combined the ratio for individual sources with the stacked
value for undetected sources weighting them by the number of sources 
(filled red squares). Open red squares indicate 
the mean value of q$_{\rm 70}$ for our SFGs {\em detected} at $70\,\um$ in the same 
redshift bins. Uncertainties are derived from measurement errors and Poisson 
statistics. The numbers beside each data point indicate the mean total IR 
luminosity (in units of $\log[L_\odot]$ and summed 
over $8-1000\,\um$) derived from the radio SFRs and the Kennicutt (1998) relation. 
The black lines again represent the tracks from local templates from Rieke et al. 
(2009) at luminosities indicted in the figure.
These lines become thinner at the redshift a given template
becomes undetectable in our radio survey. The very thin grey lines represent 
tracks from 
templates at intermediate luminosities with intervals of 0.25\,dex.
The trend of a decrease in q$_{\rm 70}$ toward higher redshifts is generally 
consistent with tracks using local SEDs. }
\label{fig:q70stacksfg}
\end{figure*}
\nocite{Kennicutt:98}

\subsection{Observed q$_{\rm 70}$ for the Stacked Sample}

From our stacked images we obtained flux densities ranging between 
$0.3$ and $2.5\,$mJy with uncertainties of $\sim0.3\,$mJy. These flux 
densities correspond to $0.15$ to $1.25$ times the rms of the original image 
($2\,$mJy) showing a typical factor of $\sim7$ improvement in sensitivity. 
For each stacked measure of the $70\,\um$ flux density we calculate
a mean value of q$_{\rm 70}$ by dividing by the mean radio flux density of 
the sources in a particular bin. In Fig.~\ref{fig:q70stackall} we plot this 
value of q$_{\rm 70}$ as a function of redshift for both individual sources and 
stacks. In the four SFG redshift bins we find significant detections of 
$70\,\um$ flux density which give values of q$_{\rm 70}$ at the lower end of the 
distribution for SFGs. Hence,
every SFG has a detection at $70\,\um$, either directly or indirectly. 
The stacked AGN are not detected in the lower ($z\le 1$) redshift bin nor in 
the unknown redshift bin, but they are detected in the highest redshift bin
($1\le z\le 5$). 
At each redshift however the stacked value of q$_{\rm 70}$ for AGN is at least 
$0.5\,$dex below that for SFGs.

\subsection{Combined Mean Flux Density Ratio}

In Fig.~\ref{fig:q70stackall} we note that the whole population of 
radio-selected SFGs including non-detections at $70\,\um$ has a scatter, 
i.e. range of values, in q$_{\rm 70}$ of at least one order of magnitude 
below $z=1$. There is also a global  trend toward lower values of q$_{\rm 70}$ 
at higher redshifts as seen with just the sources detected at $70\,\um$. 
The scatter for the AGN  below $z=1$ is larger, at least two orders of magnitude.
The track of q$_{\rm 70}$ for a local {\it radio-loud} QSO \citep{Elvis:94} 
remains very close to q$_{\rm 70}=-1$ for 
this whole redshift range (off the plot for the figures presented here). We 
note that  typical values of q$_{\rm 70}$ for radio galaxies are
$\ge -2$ using data from both low \citep{Dicken:08} and high redshift samples 
\citep{Seymour:07}.

Given that we can detect all the SFGs at $70\,\um$, either individually or 
stacked, we can then examine the mean flux density ratio of the whole 
population as a function of redshift. We decrease the size of the redshift 
bin widths to half the prior value and make new stacked $70\,\um$ images, 
but restrict our analysis to sources up to $z=3$ as we have very few SFGs 
above this redshift. This decrease in bin size is important to provide
fidelity in following any trend with redshift and is possible due to the 
larger number of sources per bin when including the detected and undetected 
sources. We combine the mean value of q$_{\rm 70}$ from the 
detected sources together with the stacked value of q$_{\rm 70}$ in these new bins 
by applying a weight scaled by the number 
of sources they represent in each bin. We derive the mean redshifts
of each bin with a similar weighting scheme in $\log(1+z)$ space. The results
of this approach are shown in Fig.~\ref{fig:q70stacksfg} which demonstrates
a steady decrease in the {\it observed} far-IR/radio ratio toward higher redshift. 
This decrease is steeper than that seen for the detected sources. 
We overlay the tracks of local SFGs
from \citet{Rieke:09} at different luminosities 
($\log[L_{IR}/L_\odot]=10, 11, 12, 13$).

\subsection{Selection Issues}
\label{sec.bias}

The work presented here, in particular Fig.~\ref{fig:q70stacksfg}, is naturally 
dependent on the initial separation of AGN and SFGs in S08 where 
a source was classified as an AGN by at least one of the methods discussed in the 
introduction. While we are confident that the selection is correct in at least 
a statistical sense it may fail on an individual basis as discussed in \S4.1. 
Several of the discrimination 
methods in S08 (e.g. radio spectral index, radio morphology) were less effective
for fainter sources, but this will change with up-coming data from surveys
by e-MERLIN and LOFAR. Such data will be particularly important as 
SFGs are expected to dominate at fainter flux densities.

One possible bias in our selection of AGN in this work is the use of the 
$24\,\um$/radio flux density ratio as a diagnostic in S08. 
We used this flux density ratio (q$_{24}$) 
to select AGN as sources that were more than $5\,\sigma$ 
away from the lowest SFG track in q$_{24}$-redshift 
space. This selection could potentially lead to a bias against SFGs with 
low values of q$_{\rm 70}$. We identify 55/179 AGN  
solely on their value of q$_{24}$ (none of which were actually detected at 
$24\,\um$ and hence are upper limits). When we repeated the whole analysis 
without the 
$24\,\um$/radio diagnostic our mean values of q$_{\rm 70}$ for the 
SFGs in Figs~\ref{fig:q70stackall} \&~\ref{fig:q70stacksfg} decreased by 
$0.03-0.1\,$dex. As this effect is small we conclude the q$_{24}$ selection 
in S08 does not significantly affect the results presented here.

As discussed in \S2.1 our initial radio survey is not complete at the 
faintest flux densities due to the effective decrease in survey area at
the faintest levels. However, we have 
previously derived and used the necessary correction factors in S04.
To asses whether their use here has any systematic effect
we reran our code without these correction 
factors and found very little difference to our results. In 
Figs~\ref{fig:q70stackall} \&~\ref{fig:q70stacksfg} we observed 
a decrease in q$_{\rm 70}$ of $\sim0.05\,$dex for the highest redshift bins
with smaller differences at lower redshifts.
Given the small size of this
change and the successful use of these correction factors in previous 
publications we believe any uncertainties in these correction factors
do not significantly affect our results.

\section{Discussion}

\subsection{Direct Detection of Radio-Selected AGN at $70\,\um$}

An interesting result in Fig.~\ref{fig:q70stackall} is the detection of eight
radio-selected AGN at  $70\,\um$. Given their redshifts, they have values of 
q$_{\rm 70}$ consistent with the SFG tracks. Hence, these sources are either 
AGN with unusually strong $70\,\um$ emission or they are SFGs mis-identified as 
AGN in S08.

Looking more closely at the reasons for their original selection as SFGs in 
S08 we see that four were selected due to their very flat or steep radio spectral 
index, three due to their compact radio morphology and one due to both its low 
radio/$24\,\um$ flux density ratio and its compact radio morphology. Of the 
sources selected by their spectral index, 2/4
have values that are close to the accepted range for SFGs. Given the low 
significance of these particular sources ($3-4\,\sigma$) at $4.8\,$GHz their 
spectral index is more uncertain, hence it is possible that these sources
may in fact be SFGs. The other two sources flagged by their spectral index have a 
very steep and a very flat spectrum well away from the range
of SFGs. The four sources selected due to their radio morphology have compact 
unresolved radio emission from a $0.3''$ beam, but extended, 
elliptical optical morphology. The radio brightness temperatures of 
these sources are not high enough to definitely confirm an AGN origin to the
radio emission and we cannot rule out the presence of compact nuclear 
starbursts. 

It is also possible that some of these sources are hybrid AGN/SFG 
objects, i.e. they have a non-negligible contribution from star formation to 
the radio (in S08 we made the assumption that one process was 
completely dominant at radio wavelengths) or the IR \citep[e.g.][]{Maiolino:95}. 
For example, a radio source with 
$30\,\%$ of its radio flux density from star formation may still be selected 
as an AGN by some of the selection criteria in S08, but the value of 
q$_{\rm 70}$ for the star formation component would be just $0.5\,$dex higher 
than the hybrid value, still within the
scatter for SFGs. 
Hence, simple flux density ratios, which currently are the dominant diagnostic
at fainter flux densities, are not capable of diagnosing such hybrid objects. 
Given the short time-scale of the starburst and radio-loud AGN phase and the 
wide range of possible radio luminosities from both these processes, it is 
reasonable to expect the chances of detecting many hybrid sources in a blind
radio survey to be low.

We note that only 8/92 sources detected at $70\,\um$ are potential hybrids 
and that two maybe SFGs. Given the statistical approach of S08, to 
have a few potential mis-identifications is not unexpected. If we were 
to assign all these sources as SFGs our results in Fig.~\ref{fig:q70stacksfg}
would change by $\le 0.02\,$dex.
The distribution of these `hybrids' in the redshift/q$_{\rm 70}$
plane is only marginally different from the other detected sources (they have
slightly lower values of q$_{\rm 70}$ at a given redshift). However, 
only two have uncertain radio AGN indicators, and the rest likely 
could be genuine AGN or hybrids as there is some evidence that Seyferts and 
radio-quiet AGN do follow the far-IR/radio correlation \citep{Roy:98}. 
This complication indicates the importance of future high resolution and 
low frequency radio data in determining the relative contribution of SFG and
AGN within a single object as well as for a flux limited sample.

\subsection{Decrease of Observed q$_{\rm 70}$ at Higher Redshifts}

In Fig.~\ref{fig:q70stacksfg} we see that at $z=0$ the q$_{\rm 70}$ values of the 
local empirical templates generally increase with total IR luminosity in the 
$10^{10-12}\,L_\odot$ luminosity range. This increase is mainly due to the 
more luminous starbursts having SEDs which peak at shorter wavelengths, i.e. 
they may be characterised as having hotter dust. This trend is seen in 
empirical observations of local starbursts \citep{Sanders:96,Rieke:09}.
The exception is in the $10^{12-13}\,L_\odot$ luminosity range where some 
of the intermediate tracks in grey have higher values of q$_{\rm 70}$ than the 
$10^{12}\,L_\odot$ tracks while the q$_{\rm 70}$ for the $10^{13}\,L_\odot$ track 
is very close to the $10^{12}\,L_\odot$ track compared to these intermediate 
tracks. In this luminosity range local sources show a 
relatively flat (i.e. $\alpha\ge-0.7$, where $S_\nu \propto \nu^\alpha$) 
radio spectrum at frequencies below $7\,$GHz which is attributed to free-free 
absorption and is reflected in the most luminous galaxy templates used here. 
While our mean values of q$_{\rm 70}$ for the SFGs follow the general decrease 
of the SED tracks toward higher redshifts they mostly follow the LIRG 
($\log[L_{\rm IR}/L_\odot]\ge11$) track at low redshifts before moving to 
values in between the LIRG and $0.1\times\,$LIRG track across 
$0.5\le z\le 1.5$. However this redshift range is where the 
inferred total IR luminosities from the radio or $70\,\um$ flux densities 
are in the high end LIRG and low end ULIRG ($\log[L_{\rm IR}/L_\odot]\sim 12$) 
regime. If our high redshift LIRGs and ULIRGs had similar SEDs to their
local counterparts we would expect values of q$_{\rm 70}$ higher by 
$\sim0.25\,$dex than we observe. 

This discrepancy between the data and the appropriate SED track by luminosity 
is a maximum of $2\,\sigma$ in the $z\sim0.86$ bin before the tracks begin to 
converge at higher redshifts. While we cannot completely rule out a few AGN 
contaminating our SFG sample we calculate that we would need at least $30\%$ 
of the 52 SFGs in the $z\sim0.86$ bin to be AGN with q$_{\rm 70}\sim 1$ to 
explain the observed difference. 

This difference in observed and expected values of q$_{\rm 70}$ implies a change 
in the SED of $z\sim1$ (U)LIRGs compared to local ones at moderate 
significance. For the ULIRGs one possibility is that there is less free-free 
absorption at high redshift which would increase their radio flux density 
relative to the IR. 
This explanation could be true if star forming regions in high redshift
ULIRGs are more extended \citep[e.g.][]{Chapman:04} and hence less optically 
thick than local ULIRGs. By extrapolating the rest-frame $\ge7\,$GHz 
unabsorbed radio power-law to the lower frequencies probed by our 
$1.4\,$GHz observations we can calculate 
the change in q$_{\rm 70}$ if no free-free absorption occurs in such objects. 
We find a decrease in the value of q$_{\rm 70}$ of 0.15(0.1)\,dex at 
$z=0.5(1.5)$. This explanation could account for a considerable amount 
of the difference in q$_{\rm 70}$ for the most luminous sources.

Another possibility is that there is a change in the IR properties of LIRGs 
and ULIRGs at high redshift. There is evidence that a significant number of 
these sources at high redshift have different SEDs in comparison to local 
analogues, although the difference is not well determined and depends on 
selection \citep[e.g.][]{RR:05,Papovich:07,Symeonidis:08}. For example, 
Symeonidis et al. (2008) present evidence that many sources selected at 
$70\,\um$ at $0.1<z<1.2$ have SEDs that are preferentially cold, while 
Papovich et al. (2008) find that the SEDs of bright $24\,\um$ sources at 
$1.5<z<2.5$, on average, are preferentially warm. It is likely that the 
structure of the luminous star forming regions are different in some 
(U)LIRGs at high redshift than locally, with more activity away from the 
compact nuclear sources that are typical for nearby ULIRGs 
\citep[e.g.][]{Chapman:04}. This explanation would also permit the changes 
in radio emission discussed in the previous paragraph. The work 
presented here is unable to distinguish between a change in the IR SED, a
change in the radio spectral index or even some combination of both.
Determining the physical reasons for the differences between local 
and distant luminous starbursts will help us to understand the rapid change 
in star formation rate density from the present day to $z\sim1$
where  these high luminosity objects dominate star forming energy budget
\citep{LeFloch:05}.
Future observations by {\it Herschel} will be able to characterise these 
differences in detail.

\subsection{Comparison with Other Studies}

Many of the recent studies of the mid and far-IR/radio correlation 
\citep[e.g.][]{Boyle:07,Beswick:08,Garn:09} have examined values of 
q$_{\rm IR}$ as function of radio and/or IR flux density. Most IR/radio 
studies have been based on $24\,\um$ data because of its higher sensitivity, 
but as we discussed in the introduction we use $70\,\um$ data here as it 
more closely traces star formation.
For our detected radio sources we found a median value of q$_{\rm 70}$ 
comparable to that of \citet{Appleton:04}, but we find evidence for a 
decrease toward 
higher redshifts unlike the non k-corrected data from Appleton et al. (2004).
We used data from a narrower, but deeper survey in the radio and mid-IR which 
could explain why we see this decrease.
In the only other study to use $70\,\um$ data Garn et al. (2009) examined the 
radio fluxes of $70\,\um$ sources by stacking and found no evolution in 
q$_{\rm 70}$ over the
$10-100\,$mJy $70\,\um$ flux range (similar to the flux range of the $70\,\um$
detected sources in this work).

\citet{Ibar:08} studied the $24\,\um$/radio correlation by examining the 
$24\,\um$ properties of a sub-mJy radio sample. They found a decrease in 
observed (i.e. when no k-correction is applied) q$_{24}$ with redshift, but 
not as steep as that seen here with q$_{\rm 70}$. The shallower 
slope is likely due to the SED of the selected sources being flatter at the
wavelengths seen by the $24\,\um$ band. The $70\,\um$ band used here probes 
longer wavelengths where the SED is rising more steeply in typical 
starbursts, hence the observed value of q$_{\rm 70}$ is likely to decrease 
more rapidly.  Using an M82 $k-$correction those authors
found an approximately constant value of q$_{24}$ up to $z\sim 3$. This value, 
q$_{24}=0.71\pm0.47$, is lower than the value of q$_{\rm 70}$ observed here at any 
redshift. This difference can be explained by the shorter wavelengths probed 
by the $24\,\um$ band which are well away from the peak of the IR luminosity 
that traces the bulk of the
star formation. The value of q$_{24}$ also depends strongly on the SED assumed 
for the $k-$correction as Ibar et al. discuss and, furthermore, this value of 
q${\rm 24}$ is based on $24\,\um$ detections only and makes no other 
discrimination between AGN and SFGs. 

For a sample of $24\,\um$ selected
sources, \citet{Boyle:07} observed a systematically higher value of q$_{24}$
than seen by other authors. This value was constant with flux density. Given
the selection wavelength and the lack of discrimination by galaxy type
these authors are potentially selecting a large number of radio 
quiet AGN which have strong observed $24\,\um$ emission, but relatively little 
radio emission. However, in a similar study of $24\,\um$ selected sources 
\citet{Beswick:08} found that the observed value of q$_{24}$ decreased toward 
fainter $24\,\um$ flux densities and also decreased slightly toward higher 
redshifts. In fact, at the faintest flux densities the \citet{Beswick:08} 
result was almost an order of magnitude lower than that seen by 
\citet{Boyle:07}. This discrepancy has not been fully explained although it 
is possible the difference is a result of the different radio telescopes 
the different resolutions used (ATCA \& MERLIN for 
Boyle et al. and Beswick et al. respectively). 

\subsection{Comparison with Models of the Far-IR/Radio Relation}

While the ultimate origin of the far-IR/radio correlation is hot, young stars 
in dusty regions of the galaxy, the origin of the radiation we observe is not 
completely co-spatial or co-temporal. The wavelength range of far-IR radiation 
probed by original {\it IRAS} $60$ and $100\,\um$ band (and MIPS $70\,\um$ 
band used here) originates from 
dust clouds relatively close to the young stars and emits fairly promptly 
after a localised burst of star formation whereas radio emission originates 
from electrons traveling 
through the galaxy and interacting with its magnetic field which can happen 
relatively far and later than the original formation of stars. This 
difference in origin of the two wavelengths explains the scatter we do see 
in the local relation \citep{Yun:01} and the scatter within galaxies 
\citep{Murphy:06}. However, this 
difference also suggests that there is a dependence of the far-IR/radio 
correlation with the nature and mode of the very recent and instantaneous star 
formation. 

\citet{Bressan:02} have performed detailed modeling of the far-IR/radio 
correlation using the concept of age-selective obscuration (i.e. where younger 
stars and their corresponding emission lie in denser molecular clouds) to  
derive SEDs and potential observables for a range of galaxy ages and SFRs. They 
find that the luminosity (far-IR or radio) to SFR ratio does vary by up to an 
order of magnitude with the age of the starburst for a range of star formation 
histories. Although these variations partly cancel out, these authors do 
predict a variation of q$_{\rm IR}$ with starburst age which appears
stronger for starbursts with shorter e-folding times. They postulate that a low 
value of q$_{IR}$ could be simply the natural consequence of a particular, e.g. 
post-starburst, phase in a galaxy's history. Hence, a further possible explanation 
for our low value of q$_{\rm 70}$ could simply be the phase the starburst is in. 
\citet{Bressan:02} do predict an evolutionary path through the q$_{IR}$/radio 
spectral index parameter space which could be used to determine the age of the
starburst should sufficiently accurate IR and radio data be available in the 
future.

\section{Conclusions}

For the first time the far-IR/radio correlation of radio-selected SFGs and AGN 
has been examined to high redshifts for a large number of sources.
The $70\,\um$ band is a more direct tracer of the bulk of the IR luminosity 
than $24\,\um$ and avoids many complicated spectral features in the mid-IR.
Throughout this paper we adopted a philosophy of studying {\em observed} flux
density ratios as a function of redshift and comparing to the ratio observed 
from a redshifted SED rather than apply a $k-$correction. This approach 
avoids the problem of choosing which template to use in making a 
$k-$correction which can be problematic given the range 
of SEDs used previously for similar work, e.g. M82 and Arp220, and their 
application at high redshift.

We have shown that:

\begin{itemize}

\item Due to current radio and far-IR survey limits only sources with 
relatively high values of observed q$_{\rm 70}$ are detected at $70\,\um$.

\item The radio sources detected at $70\,\um$ have similar observed median 
values and uncertainties in q$_{\rm 70}$ as reported previously by a similarly 
selected sample \citep{Appleton:04}, but also show evidence of a systematic 
decrease towards high redshift despite the bias towards high values of 
q$_{\rm 70}$ due to the $70\,\um$ detection limit of the currently available 
data.

\item Of those radio sources detected at $70\,\um$ 84/92 were previously 
identified as radio-selected SFGs. On closer examination due to their 
detection at $70\,\um$ two of the remaining eight sources, identified as 
radio-selected AGN in S08, are thought to be SFGs, or to show a significant 
contribution from star formation, due to uncertainty in their radio spectra.

\item When stacking the sources not detected at $70\,\um$ by galaxy type and 
redshift we find a stacked detection in each of the four redshift bins for the 
SFGs, but only detect the AGN in the highest redshift bin. In the lower 
redshift and unknown redshift AGN bins we do not find a stacked detection down 
to $\sim0.3\,$mJy. The SFGs have stacked flux densities not much below our 
original $70\,\um$ detection limit, hence the inferred values of q$_{\rm 70}$ 
are only slightly below those seen for the individually detected SFGs.

\item In order to examine the far-IR/radio correlation we determined the value 
of q$_{\rm 70}$ for all the radio selected SFGs. 
The mean values of q$_{\rm 70}$ for the detected and non-detected 
radio-selected SFGs were combined with appropriate weighting and we find an 
observed value of q$_{\rm 70}$ that decreases toward higher redshift as 
expected from tracks derived from empirical SEDs of local SFGs.

\item When the observed values of q$_{\rm 70}$ for the SFGs are compared 
closely to tracks of local SEDs at the appropriate luminosity/SFR we find 
that while they broadly agree at low redshift the observed q$_{\rm 70}$ at 
$z\sim 1$ (for sources close to ULIRG luminosities) is $2\,\sigma$ below the 
value seen for local ULIRGs shifted to this redshift. This result implies 
a difference in the SED of local and $z\sim 1$ ULIRGs, and demonstrates the 
value of future comparisons using bolometric luminosities when examining the 
far-IR/radio correlation at high redshift.
At redshifts higher than $z\sim 1$ the 
tracks of local SFGs converge and our observed values of q$_{\rm 70}$ have 
higher uncertainties, hence we are unable to determine whether such a 
difference persists to higher redshifts from this data set.

\end{itemize}

We cannot completely rule out that part of this last result is due to 
AGN contaminants, although it would require a large fraction of radio-excess 
sources.
The lower value q$_{\rm 70}$, beyond that simply inferred by 
the shifting of the band pass with redshift is consistent with other results 
in the literature that suggest that LIRGs and ULIRGs
at high redshift are a more diverse population than their local counterparts
\citep{RR:05,Sajina:06, Brand:08,Symeonidis:08}.
The full range of ULIRG properties at high redshift will be revealed by 
{\it Herschel} which will be the first telescope to truly select such 
objects based on their directly observed bolometric luminosities.

\section*{Acknowledgments}

We thank the anonymous reviewer for many helpful comments that improved the 
clarity and presentation of this paper.
This work is based in part on observations made with the {\it Spitzer Space 
Telescope}, which is operated by the Jet Propulsion Laboratory, California 
Institute of Technology under a contract with NASA. Support for this work 
was provided by NASA through an award issued by JPL/Caltech.  The National 
Radio Astronomy Observatory is a facility of the National Science Foundation 
operated under cooperative agreement by Associated Universities, Inc. 
This work was partially supported by JPL/Caltech contract 1255094 to the 
University of Arizona.

\bsp

\label{lastpage}

\end{document}